# Physics and Commonsense[*]

*— Reassessing the connection in the light of quantum theory*

Ravi Gomatam[†]

## 1.0 Introduction

In this paper, I examine the necessity of solving the philosophical problem of naïve realism at the level of everyday thinking appropriately, in order to solve the problem to scientific realism posed by quantum theory.

Broadly stated, naïve realism is the attitude that the form of our outer experiences directly and literally correspond to the structure of the real world underlying these experiences. Naïve realism permeates our everyday thinking about, and ordinary language description of, the macroscopic world. It has undeniable pragmatic justification. However, as Descartes recognized centuries ago, strictly philosophically speaking, naïve realism requires a justification. Physicists, nevertheless, *simply assume* naïve realism in interpreting the laboratory observations realistically. Thus, physicists do not find the philosophical issues surrounding the problem of naïve realism to be any of relevance to the development of physics per se. The problem of scientific realism is seen to arise in a different context, that of justifying the idea that our best theories in physics pragmatically succeed because their abstract description is a true description of the world in some sense.

Scientific realism in classical (i.e. pre-quantum) physics has remained compatible with the naïve realism of everyday thinking on the whole; whereas it has proven impossible to find any consistent way to visualize the world underlying quantum theory in terms of our pictures

---

in the everyday world. The general conclusion is that in quantum theory naïve realism, although necessary at the level of observations, fails at the microscopic level.

In this paper I offer a counter view that what fails in quantum theory is naïve realism at the level of observations itself.

Toward making the arguments clear, I introduce a detailed philosophical terminology in the first part, involving a distinction between different kinds of worlds – external, phenomenal, mathematical and physical, as well as a distinction between phenomenal and realistic modes of ordinary language (*P*-mode and *R*-mode, respectively).

The terminology is helpful to develop a detailed view of scientific realism in classical physics that I call *analogical realism*. As per this view, the theoretical objects of physics are 'real' not in the sense they can be shown to correspond to some aspects of the external world (this project is well-known to have definitively failed) but in the sense that once a theory's predictions are verified, we treat its objects as real to the extent we can visualize them using ordinary language words (and associated space/time visualizations). It is easily seen that while this realism by analogy grew increasingly tenuous with the progress in physics, in quantum theory it failed completely. We cannot consistently use either particle or wave pictures (the only object pictures in everyday thinking so far available to us). We are forced to use one or the other to visualize the one and same quantum object, albeit under different and mutually exclusive experimental arrangements. We are faced with either an ontological contradiction or a limited phenomenal conception of the quantum world that changes with changing conditions of its observation.

The terminology developed in this paper is used to diagnose where analogical realism fails in quantum theory, without entering the territory of wave-particle duality. Our laboratory observations are, first and foremost, our sense experiences. They are, and must be, described using only ordinary language. I point out that naïve realism underlying our commonsense view of the world consists of *two* parts: our outer experiences (that laboratory observations are), if veridical, pertain to a definite objective situation in the external world, and the same ordinary language sentence that describes our experience describe that reality too.

The separation between these two parts of naïve realism opens the possibility that what fails in quantum theory is not naïve realism in toto, but merely the naïve part of naïve realism. The implication is that, in order to interpret quantum theory realistically, we do not need to embrace the idea that quantum reality is essentially mathematical, or must involve a duality. Instead, we can opt to develop an *alternative* way to go from current ordinary language descriptions of our experiences to a quantum-compatible, realistic mode of describing the *macroscopic* real world. I shall conclude by arguing that the above understanding of the situation in quantum theory suggests the need to go beyond the currently prevailing Cartesian conception of matter (as *res extensa*) at the macroscopic or observational level in order to realistically interpret quantum theory.

## 2.0  Setting up the terminology

I begin by distinguishing among three types of object concepts: common-sense object concepts, mathematical object concepts, and physical object concepts. They occur respectively in common-sense thinking, mathematics, and physics.

Corresponding to these three kinds of objects, I identify three kinds of conceptual *worlds*, namely the *phenomenal* world, which is directly given in experience by definition; the *mathematical* world, which is entirely ideal; and the *physical* world (or, the world of physics, if you will), which is indirectly given in experience. Adopting a realist attitude, I additionally postulate the existence of the external world, which is different from all of these three conceptual worlds (see Figure 1).

All three object-concepts mentioned above, *qua* concepts, are equally abstract. However, common-sense object-concepts, being embodied in ordinary language, are directly coordinated to sense experiences and hence can be given an *ostensive* definition. In this sense, they are empirically the least abstract.

Mathematical object concepts, on the other hand, cannot be coordinated to experience at all, even if mathematicians often use ordinary language words such as "space" to talk of the mathematical world. For example, consider the "Hilbert space", which is the set of all functions that map from R(4) to C(2).[1] This is a very abstract 'space', and we do not expect to be able to visualize it in terms of any notion we associate with the word 'space' in common-sense thinking. Nor can the objects of this abstract Hilbert space, called the Ψ functions, be visualized in common-sense terms, i.e., be given an ostensive definition. Thus, the mathematical concepts are relatively the most abstract, again empirically speaking.

---

[1] They also satisfy a 'square-integrability' condition.

Objects and other concepts of physics — I shall refer to them as 'physical concepts' — are in-between. Being basically mathematical concepts and therefore essentially formal, they too cannot be directly coordinated to sense experiences and given an ostensive definition. Yet the physicist empirically interprets the theory in which these physical concepts play a role, by interpreting the solutions to the mathematical equations of the formalism as numbers that will appear as readings in appropriately calibrated meters. When subsequently these predictions are verified by performing actual experiments, at that point, the abstract physical concepts *can* be coordinated to sense experience. For example, although we do not expect to be able to directly point to an 'electron' in our experience, when a localized dot is interpreted as being caused by an electron, we can (and do) point to that dot and say that *this* is an observed electron. The corresponding mathematical concepts become physical concepts, at this point. For example, in quantum theory, the same Hilbert space we mentioned before is interpreted as the physical space in which the quantum dynamics takes place, and its element, the $\psi$ function, now becomes a physical state.

Note that from the viewpoint of the terminology being developed in this paper, I will *not* refer to a 'table' as a physical object. Rather, it is a phenomenal object. I reserve the term 'physical' to refer to the abstract mathematical spaces in which theoretical entities postulated by physics (physical concepts) reside. Thus, as per the viewpoint being developed here, 'physical objects' exist only in a 'physical world', which is an abstract, physico-mathematical world that is internal to each physical theory and hence could be different for different theories. The same phenomenal 'table' *qua* physical object would be treated by a 'point particle' in the physical world of classical mechanics, in terms of a world line in the physical world of theory of relativity, or in terms of a complicated wave function representing a collection of atoms in the physical world of quantum theory. The corresponding mathematics for each of these physical worlds is accordingly different too. I shall have to say more on the use of the word 'physical' in this special sense in later sections.

## 3.0 Common-sense Thinking and Naïve Realism

Einstein said, "Physics is nothing but a refinement of common-sense thinking."[2] Common-sense thinking is embodied in our ordinary language. Thus, our common-sense thinking is implicit in the way we describe our sense perceptions using ordinary language.

When we express our sense perceptions using ordinary language, we may speak of "the table in front of us" and believe that we are referring to a table that exists in the external world. This attitude is known as "naïve realism" and is characterized by the belief that the world we come to know in experience *is* the external world, as long as there are no 'errors' in perception or other mitigating factors that question the veracity of our very experience. In so describing our outer experiences, i.e., experiences of the external world, ordinary language is functioning in what I shall call the realist mode or the **R-mode** (see Figure 2).

However, ordinary language also functions in another mode, which I shall call the phenomenal mode or the **P-mode**. The ordinary language statement "I feel happy", for example, refers directly to *our inner experience* and therefore is in the *P*-mode. Similarly, when we say "the sunset is beautiful", ordinary language is functioning in the *P*-mode, since the statement refers to our inner experiential content. But, if we say "the sun is setting", ordinary language is functioning in the *R*-mode since that statement would refer to an objective state of affairs in the external world. To say ordinary language is functioning in the *P*-mode in the statement "the sunset is beautiful" is not to take a stance on the aesthetic question of whether it is objectively possible for the sunset to be beautiful. Nor am I particularly concerned by whatever implication the *P*-mode statement "the sunset is beautiful" may have for the validity of the *R*-mode of ordinary language. My intention in introducing the distinction between the *P*-mode and *R*-mode of ordinary language is not to deny that the *P*-mode statements even with regard to inner experiences could imply *R*-mode ideas (after all, we could not have a veridical inner experience of a 'beautiful' sunset if there were not actually a sunset taking place in the world), but rather to pave the way for the discussion of a *reverse* point: that the start of modern empirical philosophy is characterized by the recognition that *all* statements concerning outer experiences in the *R*-mode imply the validity of the same statements in the *P*-mode first.

---

[2] For the purposes of this paper, I shall consider only that range of common-sense thinking that is in relation to the external world (Figure 1).

## 4.0 Philosophical Realism

The start of modern Western philosophy[3] is characterized by Descartes' recognition that, logically speaking, even while describing *outer* experiences, ordinary language can be said to function *only* in the *P*-mode, and that therefore the common-sense naïve realist attitude that ordinary language functions in the *R*-mode to refer to objects in the external world is an *extrapolation* that needs *justification*. The different schools of empiricism, such as skepticism, idealism and realism (as well as any number of other schools, such as operationalism, conventionalism, foundationalism, etc.) would respond differently to the question, of whether it is possible to justify treating ordinary language as functioning in the *R*-mode in such cases.

Since physics unavoidably involves realist praxis, and my plan is to analyze in detail the connection among philosophical analysis of ordinary language realism (also called naïve realism) and physics in general, and then to relate this analysis to quantum physics in particular. The first thesis of philosophical realism is that the subject-independent existence of an external world is a metaphysical presupposition, since the external world is *not* directly knowable as such. Given this presupposition, the next nmmmj is to postulate that the causal interaction of our senses with the objects of the external world gives rise not to sense perceptions but to semantically neutral sense impressions, which are then described by ordinary language (see Figure 3).

The distinction I draw between sense impressions and sense perceptions can be explicated as follows. In common-sense realism, if we report seeing a table in front of us, we assume that if we had merely looked, without saying anything, we still would be seeing a table. That is to say, the mere act of looking, i.e., letting our eyes causally interact with the world, results in *perception*, which already has cognitive meaning exactly the same as that of a subsequent ordinary language report ("I see a table in front of me"). The language simply enables us to ransparently report what we see.

---

[3] I restrict myself in this paper only to a consideration of the analytical tradition of Western philosophy, and in particular to a sub-field called "ordinary language philosophy". Similarly, in discussing science in this paper, I shall restrict myself to physics.

However, within philosophical realism as I espouse here, what we register when we merely look are "sense impressions." They do not have *meaning*, i.e., they are semantically neutral. In this sense, the interaction of our eyes with the world is no different from the interaction of a camera lens with an object. In either case, there is no cognitive act of perception yet.[4] However, when we do describe the sense impressions using ordinary language, the act of cognition is certainly completed, since we always talk of an actual experience. If so, somewhere between when the sense impressions are produced and when we state in ordinary language our experience, our act of cognition is consummated.

But where and how does this happen? Evidently, the problem of how we come to have an experience such as "seeing a table" is very important for cognitive science.[5] However, ordinary language philosophers are concerned with another problem, namely that of analyzing *what it means to say* that we are seeing a table. In other words, ordinary language philosophers see the philosophical analysis of the problem of perception as an analysis of the common-sense intuitions behind the corresponding ordinary language statements. Although this may at first seem counterintuitive and an infertile exercise, we shall see later that it is not without its relevance because of the critical role that ordinary language plays in physics too (even though ordinary language philosophy developed independent of physics).

The immediate point is that, in describing our experience, ordinary language is functioning in the *P*-mode, and therefore the 'objects' referred to in our ordinary language statements, in the first instance, are not objects in the external world, but are only given in our experience. Indeed, philosophers refer to the objects given in our experience as phenomenal objects and take them to exist in the phenomenal world. Only via an act of *further interpretation*, philosophical realists would hold, do we transit from the *P*-mode to the *R*-mode to view our experiences as direct experiences *of* the external world.

---

[4] Obviously, this point applies to our other four sense modalities as well.

[5] While it is most likely correct to hold that semantically-neutral sense impressions are all we have to start with, recent experiments in psychology of perception seem to show that stages may be involved in perception between the formation of sense impressions and eventual ordinary language description of experience. Such a conclusion is also in line with the proposal I make in the concluding section, that Cartesian dualism must be extended by admitting more types of matter than merely the extended substance.

In philosophical realism as I present it here, there exists an implicit distinction between the "real" world and the external world. The external world exists independent of us, in and of itself. The real world is what can be said to "exist" when we realistically interpret our experiences using ordinary language in *R*-mode. Of course, the philosophical justification of the thesis of common-sense realism must eventually accept that the structure of the real world as given in our experience will correspond to the structure of the external world, but it is useful nevertheless to make the above distinction between real world and external world in philosophical realism.

I define philosophical realism, then, as the project to justify the transition from the *P*-mode to the *R*-mode of ordinary language usage when describing our simple sense experiences. This project is already counterintuitive, since common-sense thinking takes the validity of this transition for granted. Moreover, even after centuries of effort, philosophy has failed to logically justify naïve realism. As a result, it is an open secret that this part of modern analytical philosophy, which sees certain problems in the use of ordinary language where common-sense usage sees none, has failed to be intuitively meaningful and accessible to the general public.

And it is precisely for the same reason that empirical philosophical realism has also failed to excite the imagination of the physicist, since physicists simply *assume*, both in the classical and quantum domains, the validity of the *R*-mode of ordinary language in describing the laboratory events in moving toward scientific realism. By the same token, if it ever turns out that physicists are unable to make this assumption, and instead find it necessary to justify the transition from *P*-mode to *R*-mode of ordinary language with regard to our simple sense experiences within physics, then philosophical realism would become relevant to science. Since this is a crucial point to the theme of this paper, we shall investigate scientific realism in some detail and contrast it with everyday and philosophical realism in the next few sections.

## 5.0 Scientific Realism

Scientific realism holds that empirically correct theories in physics succeed because they provide a true description of the relevant part of the external world. This naïve view of scientific realism has its origins in classical mechanics and is thus very compatible with it.

A quick account of the details of this naïve view of scientific realism might run as follows: the scientists create a model of the external world out of their own imagination to account for sense phenomena, yet, unlike other forms of explanation — philosophical or religious, for instance — scientific models can be proven by testable empirical consequences. A scientific theory is accepted if and only if all its testable predictions are shown to be true through appropriate experiments. Once a theory is shown to be empirically true and technological applications are made based on its correct predictions, it follows that its description of the underlying world must be true. How else could the predictions be true and the machines work? Thus, science can be taken to construct a *true model* of the external world. This model, of course, can change from theory to theory. We might even replace an older scientific model by a newer one — for strictly empirical and hence demonstrable reasons that scientists have worked out — but this supplanting need not mean that present models are passing fictions. Rather, science builds a true model of the external world by a series of approximations, *each* of which has demonstrable empirical validity and truth content.

The key conviction underlying this naive view of scientific realism is that if a theory in physics makes a correct prediction, then the 'state' of the system at the point of observation is a real state. That is, it is a state that existed in nature prior to and independent of our observation. Indeed because the state is real, the theory makes the right prediction. In this view, our observation just passively reveals a state preexisting in nature. Quantum theory, however, startlingly unseats this classical intuition and indicates that the *opposite* is true: namely, the physical theory predictively succeeds only because the state of the observed system *could not* have existed in nature prior to the actual measurement.[6] In other words, quantum theory seems to force us to distinguish between the ontological state and the observed state of a system (and hence the external world and the observed world). What we observe is a state that we can attribute to the system only in the context of that observation.

---

[6] See Appendix A for an explanation of the technical basis for this statement.

This is the origin of the usual, popular statement that in quantum theory "our observation creates the reality we observe."

Does quantum theory say anything about the unobservable real state of the world? Yes. This state is the "superposed" state. However, an observation upon this state will *always* produce an "eigenstate" that is different from this superposed state. Any connection we might wish to make between the unobservable real (i.e., superposed) state and the observed eigenstate (say, how does the eigenstate appear from the superposed state?) *must be justified*.[7]

Quite apart from the deep problems that this situation has posed in quantum theory, we can at once notice the similarity of the form of the above implication of quantum theory and the contention of empirical philosophy, namely, that what we can know is only an observed or phenomenal world, which is different from the real world and that any connection we make between the two requires justification.

Indeed, since the inception of quantum theory (nearly a century ago, with the invention of the quantum postulate by Planck to account for the observed 'blackbody radiation'), it has been recognized that quantum theory has irrevocably invalidated our classical (i.e., pre-quantum) notions of scientific realism. As a result, there has been a search for a proper understanding of the non-classical view of the world given by quantum theory. In order to truly understand how quantum theory is non-classical, we may do well to first properly comprehend the elements of the scientific realism in the classical era in greater detail, so we can accurately judge which part of this classical realism has failed in quantum theory.

Accordingly, I now furnish an account of the realist praxis in classical physics in terms of a 'scientific method' (see Figure 4). It is not my claim that physicists actually and always follow this 'method' in the stated sequence. As Einstein famously (and I believe correctly) stated, the physicist is a "methodological opportunist". The physicist does not stick to one epistemological perspective (even that of realism) at all points of his or her scientific inquiry. For the physicist, empirical correctness is the first major criterion, and whatever route takes him there is the one he must take.

---

[7] Justifying this connection is the essence of the famous, as yet unresolved, "measurement problem" in quantum mechanics.

> [N]o sooner has the epistemologist, who is seeking a clear system, fought his way through to such a system, than he is inclined to interpret the thought-content of science in the sense of his system and to reject whatever does not fit into his system. The scientist, however, cannot afford to carry his striving for epistemological systematic that far. He accepts gratefully the epistemological conceptual analysis; but the external conditions, which are set for him by the facts of experience, do not permit him to let himself be too much restricted in the construction of his conceptual world by the adherence to an epistemological system. He therefore must appear to the epistemologist as a type of unscrupulous opportunist: He appears as *realist* insofar as he seeks to describe a world independent of acts of perception; as *idealist* insofar as he looks upon the concepts and theories as the free inventions of human spirit (not logically derivable from what is empirically given); as *positivist* insofar as he considers his concepts and theories justified *only* to the extent which they furnish a logical representation of relations among sensory experiences. He may even appear as *Platonist* or *Pythagorean* insofar as he considers the view point of logical simplicity as an indispensable and effective tool of his research.
> [Einstein, 1949, p. 684, italics in the original]

Thus, I offer the following construction of the scientific method only for the limited purpose of aiding the task at hand, namely an analysis of the connections among physics, philosophy and common-sense thinking. I do believe, however, that the method portrayed here accurately takes into account the diverse elements of the praxis of physicists in general.

The conception of the 'scientific method' illustrated schematically in Figure 4 consists of the following steps:

Given a problem, say accounting for certain data, the physicist begins with certain intuitions about the underlying causal processes in the external world. This intuition is *direct* in some sense.[8]

Based on these intuitions, the physicist uses mathematical concepts to "freely create" (Einstein's phrase) "physical concepts" that are different from the original mathematical concepts because they now express the intuitions of the physicist.[9]

---

[8] "The cornerstone of science's own structure [is] the direct perception by consciousness of the existence of external reality….Now that is something which the skeptic questions in regard to religion; but it is the same in regard to science." [Planck, 1931, p. 218]

[9] Sometimes, the physicist even creates the needed new mathematics first. Newton's invention of calculus is the first and famous example of it. Heisenberg too created matrices (although mathematicians had already invented matrices; Heisenberg simply didn't know about them). In modern times, the work of Witten and others in string

The physicist then develops hypotheses that govern relations among these concepts.

Based on these hypotheses, and via complicated steps of physical and empirical reasoning, the physicist develops a physical formalism (with mathematical equations) whose solutions are real numbers which will appear as pointer readings on appropriately calibrated meters as a result of a laboratory procedure called measurement, involving *both* state preparation and observation. These laboratory events — pointer readings, detector clicks, magnetically recorded data, etc. — are the predictive content of the physical theory, which are expressed differently in different stages.

**Stage 1:** The above predictive content is expressed using ordinary language in the *P*-mode. The laboratory events now are our simple sense experiences occurring in the *phenomenal* world (Machian sensationism).

If the Stage 1 predictions are found to be correct (by performing actual experiments), then the theory is empirically verified. One now adopts further interpretive steps to move toward scientific realism.

**Stage 2:** The predictive content of the physical theory is now re-expressed using ordinary language *assuming* the validity of the *R*-mode that drives common-sense thinking, to interpret the predictions of Stage 1 as objective events in the real world.[10]

Thus, an observation statement of the form "the meter needle is pointing to +1", which would be a statement about *our* experience at Stage 1, would now be a statement about the state of a real meter in the real world which existed in the world *prior* to our observation. Evidently, the observations now take place in the *real world* of everyday thinking.

**Stage 3:** The predictive content of the physical theory is now restated using the formal terms of the physical theory (forming the vocabulary of the theoretical language). The observations

---

theory also have led to much advancement in pure mathematics. As to why mathematical concepts at all are used in physics, we shall merely take their use in current praxis as a fact. We shall not investigate what Wigner called "the unreasonable effectiveness of mathematics in physics," although the ideas in this paper do contain one possible answer to this question.

[10] We are separating Stage 1 and Stage 2, both to account for Mach's essentially correct basic insight, and to enable us to see in a later section how quantum theory may require, precisely at this Stage 2, a non-classical move.

of Stage 2 are now *scientific* observations. Thus, the observation "meter needle is pointing to +1" is now a measurement of the value of a property, such as "the property corresponding to observable x has been measured to have an eigenvalue of +1" at the point of this observation.

These scientific observations take place evidently in the abstract, *formal world of physics*. This 'physical world' would be different for different physical theories. For example, the dynamics of Newtonian mechanics takes place in the physical world housed in the abstract, mathematical Euclidean space. The physical world of relativity theory is the abstract, mathematical Riemannian 4-dimensional manifold. The physical world of quantum theory is the abstract, complex, Hilbert space.

**Stage 4:** The motivation to treat the abstract physical world as 'real' is supplied by *naming* the various formal terms using words of ordinary language in *R*-mode. The scientific observation of Stage 3 is now restated to give it a realistic flavor. Thus, the observation of Stage 3 would now be restated as "the electron was observed to be at the top slit" (corresponding to the value +1 on the meter needle), or that "the electron's spin was measured to be in the "up" direction along the z-axis."

**Stage 5:** Because the physical world is shown to be 'real', the value measured for the physical property under observation at step 8 would now be treated as having belonged to the observed system *prior* to the observation.[11]

For now, the important point is that the restating of the scientific observations (initially stated using ordinary language in *P*-mode) using the theoretical terms *named using ordinary language words* enables the physicists to communicate the felt 'reality' of the abstract, non-sensible, physical world in terms of our only source of knowledge of the real, namely the common-sense notions of the real world of everyday experience. The ordinary language words play only a suggestive or evocative role, since the identity between the physical object and the common-sense object corresponding to its name is never perfect, only an *idealization*. Therefore, though the ordinary language words enable one to visualize the formal physical world in terms of common-sense space and time pictures, one accords the mathematical formalism itself the primacy in determining the true meaning of these terms. Thus, we conclude that the physical world is rendered real only by analogy to the real world of common-sense thinking. This analogical mode of visualization is needed because the real objects of common-sense thinking are directly coordinated to sense experience, whereas it is the physical theory as a whole that is directly coordinated to experience, and the physical objects which are part of the theory are coordinated to our sense experiences only indirectly. I shall call this *analogical scientific realism*.

---

[11] The reader would recognize that we had previously mentioned that this assumption holds in classical physics, but not in quantum physics. However, our detailed analysis can aid us to check whether some other classical prejudice fails antecedently. Indeed, later we will be proposing that the antecedent classical prejudice to fail in quantum theory is in step 7.

If the physical world is entirely conceptual, how are we to understand the physicist speaking of an electron interacting with a *macroscopic detector*, since the latter is an object in the lived world? We propose that any such reference the physicist makes is not to a particular (i.e., this or that) detector, but any detector in the *generic* sense. Thus, the macroscopic detectors that inhabit the physical world (i.e., the world of physics, in which the physicist analyzes and interprets the experiment) are not the common-sense objects (i.e., the specific detectors that the physicists handle in the laboratory), but the corresponding *universals*. This renders the physical world fully conceptual, despite housing both formal entities such as particles or waves and laboratory equipment (since these are all now universals). In this way we can understand that despite the physicists' references to abstract physical entities interacting with detectors, the whole account of the experiment by the physicist refers to happenings in a conceptual physical world, *not* the real or external world. As we shall presently see, this description meets the happenings in the real world only at the interface of *our* experiences. In other words, the physicist himself provides the over-arching connection between the physical world and the world of observations.

We can now see why the physicist, while analyzing the experiments as performed in the 'real' world, i.e., the laboratory, nevertheless is able to interpret the *same* laboratory event *qua* our simple sense experience as *different* physical events under different theories. The same sense experience, say a meter needle pointing to +1, would be interpreted by the physicist in one theory as (say) the current being measured as one ampere by the meter, or the spin of an electron being 'up' in another theory. This possibility exists because, while the observation in Stage 1 (of Figure 4) is a sensible event in the phenomenal world, the interpreted scientific observation is an event occurring in the physical world.[12]

---

[12] Thus we can understand Einstein's reported statement, "it is the theory which decides what is observable."

The distinction between the phenomenal and the physical becomes even more explicit when we consider that the same physical theory could allow us to make two measurements corresponding to two different properties (involving two different experimental arrangements) and come up with the *same* sense experience (i.e., a laboratory event such as "a meter needle pointing to +1"). This possibility ordinarily does not exist in common-sense thinking, treated as a theory.[13]

To summarize, physics in particular and science in general constructs only an abstract world, which is a conceptual grid that helps us to impose an order upon a certain range of sense experiences, and put this order to practical use. Scientific realism is the claim that this conceptual world is every bit as 'real' as the external world by drawing analogies via ordinary language between the two worlds. To say the physical world is 'real' is not to say it is a model of the external world, but that we can visualize, *within physics*, its abstract, conceptual 'physical world' using ordinary language terms in much the same way we use them to visualize the external world that we regard as "real" in common-sense thinking.

## 6.0 Empirical Science and Common-sense Thinking

We have suggested that common-sense thinking embodied in ordinary language enters physics in two stages:

To describe the laboratory events *qua* our simple sense experiences
(Stage 2 in Figure 4)

To name the theoretical terms using ordinary language words to understand the physical content of the theory (Stage 4 in Figure 4)

Despite the growing abstraction of the formal physical theories, the possibility for making the connection between physical theory and common-sense thinking at Stage 4 remains crucial for understanding the physical theories realistically.

---

[13] Kant, for example, did not see the need to distinguish between the phenomenal and physical, because science of his times did not need this distinction, though it was compatible with it. My whole purpose in drawing out this distinction is to show, in subsequent sections, that the changes wrought to our physical understanding by quantum theory could be grasped better if we do not equate the phenomenal and physical even in understanding the pre-quantum or classical theories.

Thus, if it is true that science progresses by often superceding our common-sense thinking at Stage 2, it is equally true that science cannot but return to make connection with common-sense language at Stage 4 to ground scientific realism. In Stage 2, our common-sense analysis of the laboratory events might suggest one picture of the processes in the external world[14] underlying the observed phenomena, while the scientific account of the same phenomena in terms of the processes in the *physical world* might be different. For example, from the observations of the sun's relative position in the sky, we might conclude in our common-sense thinking that in the external world, the sun is rotating around the earth. Now, in the Copernican model the same observations of the relative positions of the sun at various times of day are predicted by constructing underlying processes in the *physical* (i.e., conceptual) *world* in which the 'earth' is presumed to go around the sun. But in the physical model developed in Stage 3, we have only two point-particles (theoretical constructs) in relative motion, though *we name* one as 'earth' and the other as 'sun'. Thus, the earth and the sun *qua physical objects*, whose motions we use to explain the observations, exist only in the conceptual physical world, and are different from the earth and the sun qua objects of the *external world* that we associate with the *same* observations in common-sense thinking. At this stage of theory-making, or in other words, pragmatically successful scientific-myth-making, we can adapt the quip Russell used to emphasize the difference between phenomenal and external world ("the sun I see is not the sun I see") and say "the sun we see is not the sun of the solar model." The sun around which the earth is taken to rotate (in the physical world) is not the sun that is taken to rotate around the earth (in the external world).

To give another example, let us say our *observations* are of the relative positions of "a steel ball rolling down an inclined plane". This statement concerns, in the first instance, the phenomenal steel ball. However, in common-sense thinking, we are inclined to assume that there is also a 'real' steel ball in the external world accounting for the perceived motions. Now, the same observations involving the phenomenal steel ball are predicted in Newtonian mechanics by postulating the motion of a point-particle that we would also name as 'steel ball'. Yet the point-particle that is *named* 'steel ball', whose motion in an abstract, Euclidean space classical mechanics describes, is an object of the physical world — whereas the real steel ball that common-sense speaks of is an object in the external world. In Newtonian mechanics, the point-particle 'steel ball' is usually taken to be a model of, and thus *correspond to* the real 'steel ball' in the real world.

---

[14] In this section, I do not differentiate between the real and the external world while discussing common-sense notions, since the distinction is not significant for the purposes of this section.

Despite the success of this soothing correspondence realism, even classical mechanics *need not* be seen as explaining the motion of the steel ball in the real world. We saw that empirical philosophy makes a distinction between the *P*-mode and the *R*-mode of ordinary language. Thus, a statement of the type "the steel ball is rolling down an inclined plane" would be regarded in the first instance as a *P*-mode statement in philosophy, which would connote an infinite succession of *our individual experiences,* each of which could be described as an individual experience of the sort "a steel ball at this location". Only by the subsequent assumption of naïve realism can we be said to regard the statement "the steel ball is at this location" as a statement about the location of a real steel ball in the real world, and the statement "the steel ball is rolling down an inclined plane" as a statement about the real motion of a real steel ball in the real world.

In tracing in detail the stages of scientific realist thinking, we too have viewed the laboratory events as described by ordinary language in the *P*-mode in Stage 1. If we *stop* with this viewpoint, we get the Machian sensationism in which the theoretical language of the formalism is regarded as reporting laboratory events *qua our experiences*, and the physical law is no more than a shorthand way of connoting the succession of our experiences that arise when we carry out well-defined laboratory procedures involving both state-preparation and observation. But it is not necessary to stop with this. The mistake of Mach—one that all the great system builders in the Western philosophical tradition have made — is to take a single essentially correct insight within a domain of study and use it to *complete* the understanding of the whole domain. Mach made this error with respect to science. Thus Einstein, who acknowledged that Mach's *History of Mechanics* had "exercised a profound influence upon me while I was a student", and that Mach's "epistemological position influenced me very greatly"; nonetheless, in his later years, emphasized that Mach's skepticism is essentially untenable. "Mach's system studies the existing relations between the givens of experience; for Mach, science is the totality of these relations. That point of view is horribly wrong; in short, what Mach has made is a catalogue, not a system." [Einstein quoted in Howard, 1993, p. 216] What I am doing in this part of the paper is to recognize Mach's essentially correct insight, namely that we must begin with the position that involves minimum assumptions, but *add* to that position, consciously recognizing the assumptions we add at each stage, to reach a realistic view of physics in general that will show in what manner such realism fails within quantum theory.

The potential advantage in recognizing the steps involved in scientific realism in such a detailed manner as above, even in the era of classical physics, is this: if ever our current range of naïve realist notions turns out to be inappropriate for a physical theory at *the level of observation itself*, (as I shall argue is the case with quantum theory), then instead of concluding that scientific realism has failed (as is the current fashion in interpreting quantum theory), we can examine whether there is *another route* to go from the *P*-mode to the *R*-mode in ordinary language to re-describe the laboratory events, a route which would be more quantum-compatible.

## 7.0 More on Scientific Realism

With respect to the phenomena accounted for by early mechanics, common-sense thinking directly presents the objective processes underlying these phenomena, since the two are identical. The experience of a (phenomenal) 'apple falling down' directly suggests the underlying objective process in the real world, namely that of a real apple falling down in the real world. However, with progress in science our observations (which are simple meter readings) no longer directly suggest in common-sense terms *a picture* of the objective processes in the external world underlying these observations. For example, moving a coil within range of a magnet induces a movement in the needle of an ammeter appropriately connected. This does not directly suggest the objective process underlying the observation. The physical explanation of such observations in electromagnetic theory involves positing a 'field' (an abstract, conceptual entity). Physicists visualize this 'field' as a 'wave' (an ordinary language word with an associated picture) in an approximate or idealized manner.[15] The logical force of the scientific realism, in treating the field as a 'real' object within electromagnetic theory, comes then from the fact that we invoke our only notions of the 'real', i.e., the ones we have in common sense ('wave' in this case), to visualize the physical ideas by *analogy*. The physicists cannot logically demonstrate any correspondence between such visualization and the relevant causal processes in the external world, since we sense nothing about these causal processes beyond the meter readings. However, this absence of a logically demonstrable correspondence between the external world and the physical world is true even of early mechanics, strictly speaking. Newton himself recognized and emphasized this absence.

---

[15] For example, a wave in common-sense thinking would always have a medium, i.e. be a wave of something, whereas the electromagnetic 'wave' is not a wave of any medium.

> I use the word 'attraction', 'impulse', or 'propensity' of any sort toward a centre, promiscuously and indifferently, one for another, considering those forces not physically but mathematically; whereof the reader is not to imagine that by those words I anywhere take upon me to define the kind or the manner of any action, the causes or the physical reason thereof, or that I attribute forces in a true and physical sense to certain centres (which are only mathematical points) when at any time I happen to speak of centres as attracting or as endued with attractive powers.
> [Newton, 1687, Definition VIII, pp. 5-6]

With further growth of physics, we are even guaranteed that there *cannot be a direct correspondence* between the external world and the physical world, as in relativity theory. Here too, we have simple phenomenal experiences of observed bending of the light from a star in the presence of the sun, a body of great mass, but no clue within common-sense thinking about the real processes taking place in the presumed space *and* time of the external world. The physical theory, however, explains it in terms of a physical world that is a 4-dimensional mathematical *continuum*. We are therefore certain that the physical world now is a total abstraction, even though it predicts the same experiences that we have in the phenomenal world, namely the bending of the light.

Even so, the most abstract 'physical' 4-dimensional continuum has been given a realistic flavor by naming it as a 4-dimensional *space-time* continuum. This is made possible by the consideration that if we cut the 4-dimensional continuum along a chosen plane (representing time), the remaining 3-dimensional surface can be regarded as space. Thus, an actual observation event could be regarded as taking place in space *and* time, even within relativity theory. Thus, to view the mathematical 4-dimensional Riemannian manifold as a 'space-time continuum' has some, even if only very tenuous, justification. It allows one to regard the physical world of relativity theory as 'real' in the same way we regard the common-sense world of space and time as real, simply because the same words are used to name both. Nevertheless, the move to name the freely-created theoretical terms using ordinary language words evokes a sense of the real, *not* actual correspondence. This is true of physics at all stages of its development so far. Thus, Einstein could say, "I believe physics concerns the 'real', but I am not a realist".[16]

The need for compatibility with common-sense pictures of the real world marks the distinction between strictly mathematical theories and theories in physics which use mathematics. Pauli summed it up quite eloquently.

> It is true that these [scientific] laws and our ideas of reality which they presuppose are getting more and more abstract. But for a professional, it is useful to be reminded that behind the technical and mathematical form of the thoughts underlying the laws of nature, there remains always the layer of everyday life with its ordinary language. Science is a systematic refinement of the concepts of everyday life revealing a deeper and, as we shall see, not directly visible reality behind the everyday reality of colored, noisy things. But it should not be forgotten either that this deeper reality would cease to be an object of physics, different from the objects of pure mathematics and pure speculation, *if its links with the realities of everyday life were entirely disconnected.*
> [Pauli, 1994, p. 28, italics mine]

Our proposal that a physical world is to be viewed as 'real' not solely by virtue of its predictive correctness, but also by demonstrating its visualizability (in terms of our notions of the real world in common-sense thinking) has the consequence that, if a theory were to be empirically successful but not properly amenable to visualization using ordinary language terms, physicists should be expected to regard our 'understanding' of the physical content of a theory to be incomplete. That this indeed is the case is best shown by quantum theory. On the one hand, quantum theory is predictively the most successful theory we have yet of the relevant phenomena. Yet its formal terms resist a neat, single picture of the world in ordinary language terms. We are forced to use mutually exclusive wave-particle pictures to visualize the physical picture underlying even a *single* observation.[17] Because such a visualization of quantum theory's formal terms is not fully self-consistent, despite its prodigious empirical success, scientists do regard our efforts to 'understand' quantum theory as incomplete.

---

[16] If the physical world can be said to be 'real' only in the limited sense of being capable of visualization by analogy to the real external world, and physical theories, despite their demonstrable pragmatic value in the external world, would fail to aid in our understanding of it, would such a limited scientific realism be satisfactory? Wouldn't a demonstration that science can also contribute to our increased understanding of the real, external world give a fuller credence to the stance of scientific realism, as well as more adequately respond to the social constructionist's critique of science? I believe the answer is yes, if one adopts a viewpoint called 'tandem realism,' in which common-sense thinking is expected to evolve, along with scientific thinking. I explore tandem realism in a forthcoming paper.

The foregoing discussions should explain our focus on the analytical philosophical tradition in the West, and particularly ordinary language philosophy, to discuss the connection between philosophy and science. In science, ordinary language considerations are essential.

## 8.0  Science and Philosophy — The Missing Connection

We can now state why the connection between empirical philosophy and empirical science has not so far taken place, in physics.

To recapitulate, philosophical realism in ordinary language philosophy concerns justifying the transition from the *P*-mode to the *R*-mode in common-sense thinking. We have proposed that physics too, in its first stage, predicts the observations *qua* phenomenal experiences. Thus, an instrumental view of the physical theory has been taken to be the logically correct starting point for physics, just as, for empirical philosophy, a necessary starting point is to distinguish between the phenomenal world and the real world (or between the *P*-mode and the *R*-mode use of ordinary language). However, philosophy has struggled to justify the transition from *P*-mode to *R*-mode in common-sense thinking, whereas physics has *assumed* the validity of the *R*-mode of ordinary language when reporting the laboratory events in Stage 2 (Figure 4). Scientific realism wrestles with the *subsequent* step of justifying the transition from the *P*-mode of the *theoretical* language of formalism, which is used to predict the observations *qua* sense experiences (Stage 3), to the *R*-mode in treating the same observations as 'real' physical events, by naming the theoretical terms using ordinary language words (Stage 4). For this reason, the concerns of philosophical realism and scientific realism have been entirely different.

---

[17] The popular idea that the individual quantum called 'electron' or 'photon' (with respect to matter or radiation) appears as a particle in some experiments and as a wave in some other experiments is quite superficial. It is more accurate to say that in order to account for *all* aspects of even a single observation (both why a localized event takes place and why it takes place where it does) requires invoking *both* particle and wave pictures.

This difference, not often clearly stated in the literature, can serve to explain, though not altogether justify, why although Descartes' philosophy served to trigger the start of modern Western science, physicists have come to regard philosophy as largely irrelevant for their realistic concerns. Philosophy conducts a *logical* analysis of experience described using ordinary language in relation to the *external world*. Physics conducts an *ontological* analysis of the same experience described using the theoretical language in relation to the *physical world*.

One major response to alleviate this irrelevance of empirical philosophy to the concerns of empirical science came from the logical positivists in the early part of the twentieth century, who proposed to redo philosophy by letting philosophy *follow* science. Reichenbach, for example, contended that philosophy must be done in the context of theory justification in science, i.e., *after* a theory is invented by physicists, empirically tested and accepted by the scientists, not before or independent of science. Quine was to say more succinctly, "philosophy of science is philosophy enough."

However, the positivist response is only one possible response. Significantly, it is even inappropriate in the context of one major theory in science: quantum theory. I shall now argue that a full solution of the outstanding problem of understanding the physical meaning of quantum theory at the level of the quantum world is centrally connected with the proper solution of the philosophical problem of the relation between the *R*-mode and *P*-mode of ordinary language at the level of the macroscopic world. If so, philosophy would have to operate in the context of theory creation, not just theory justification, and 'doing philosophy' would become an essential part of 'doing science.'

## 9.0 Quantum Theory and Human Experience

We have noted that a physical theory is connected to experience via several stages. We have also repeatedly stressed that the physical world, though essentially mathematical and therefore ideal, is rendered 'real' by evoking the image of the real world of common-sense thinking by naming the terms of the physical formalism using ordinary language words.

In quantum mechanics we saw that we have had to pay a much higher price to make such a connection between the formal terms and ordinary-language-based, common-sense thinking. We are forced to employ a wave-particle duality. At this level, the problem would seem to be a conflict between the ontological quantum physical picture and common-sense thinking. Indeed, this is how quantum theory is invariably presented. True, this is acute enough. However, I shall presently argue that the real problem in quantum theory may well be far more acute. In quantum theory, ordinary language can function *only* in the *P*-mode while describing the observations, and thus the usual assumption of the validity of the classical *R*-mode of ordinary language *even at the level of the observations* breaks down.

Consider the standard two-slit experiment. The electron can go through one of the two holes and thus have one of *two* possible 'positions' if observed close to the two-slit screen. Indeed, if we place two detectors close to each slit, we find that only one of them always clicks. However, either observation should not be regarded as revealing the position of the electron in the pre-quantum or classical sense, since in a three-slit experiment, the same electron in the same prepared quantum state would have one of *three* positions. In other words, even so simple a property as that corresponding to the x-observable in quantum mechanics could not be regarded in any straightforward manner as a 'position' in the classical sense of an absolute location in space, *even at the point of an actual observation*, since the observed position is due as much to the details of the experimental arrangement under which the quantum object is observed (the number of slits in the screen, in this case) as due to the electron itself.

Bohr is well-known for emphasizing this quantum inseparability.[18]

---

[18] It is true that even in classical electromagnetic theory when the wave passes through the N-slit screen, the number of wavelets that would emerge on the other side of the screen is N. Our discussion pertains, however, to the non-objective character of the quantum physical state, as compared to the classical state. In the case of classical theory, the physical state of the electromagnetic field interacting with the N-slit screen remains the same regardless of the value of N we choose for the N-slit screen; whereas, in quantum theory, the physical state of the electron or the photon interacting with the screen changes with the value of N. For example, in a 2-slit experiment, the superposed state of the electron would be $\Psi = \alpha_1 \psi_1 + \alpha_2 \psi_2$. In the case of a 3-slit experiment, the *same* electron would now be in the

> The finite magnitude of the quantum of action prevents altogether a sharp distinction being made between a phenomenon and the agency by which it is observed, a distinction which underlies the customary concept of observation and, therefore, forms the basis of the classical ideas of motion. [Bohr, 1934, p. 11-12]

> In any interaction, *implied by the quantum*, the measuring instruments and the atomic objects are inseparably entailed in the phenomena. [Bohr, 1957, p. 89-90]

However, we must inquire as to what exactly is the implication of this epistemic inseparability for our understanding of the nature of quantum *physical* properties.

> The properties of position and momentum are not only incompletely defined and opposing potentialities, but also in a very accurate description, they cannot be regarded as belonging to the electron alone; for the realization of these potentialities depends just as much on the systems with which it interacts as on the electron itself. [Bohm, 1951, p. 620]

In the above passage, Bohm speaks of the mutual contribution of the observed system and observing agency to *actualize* the measured value. This is indeed the orthodox view. However, we have seen above that in a more crucial sense, the experimental arrangement contributes to even *defining* the very *meaning* of a quantum mechanical property, even one apparently so simple in conception as 'position'. Thus, one could say that the observed system and the observing arrangement together weave a single inseparable *epistemic whole* at the point of an observation. It is a great merit of Bohr that he recognized and emphasized that quantum inseparability concerns the very definition of what it means to use the word 'property' within quantum mechanics.

---

superposed state, $\Psi = \alpha_1\psi_1 + \alpha_2\psi_2 + \alpha_3\psi_3$. These two superposed states are *physically different* due to N being different. The same is true with regard to 'position' too. Thus, the measurement problem involves, not only why the superposed position-state changes into a definite position-state in an actual observation, but also why the ontological superposed state itself changes depending on the value *we choose* for N in the N-slit screen. Our aim is to suggest that this difficult problem of measurement can perhaps be altogether avoided by reasoning that the quantum inseparability precludes the classical meaning we associate with the idea of 'position' at present (see also Gomatam, 2003a).

> Quantum mechanics speaks neither of particles the positions and velocities of which exist but cannot be accurately observed, nor of particles with indefinite positions and velocities. Rather, it speaks of *experimental arrangements in the description of which* the expressions 'position of a particle' and 'velocity of a particle' can never be employed simultaneously.
>
> [Neils Bohr, *Second International Congress for the Unity of Science*, Copenhagen, June 21-26, 1936, emphasis mine]

If the electron and the experimental arrangement form a single epistemic whole, such that we cannot use ordinary language to even describe the laboratory events as referring to the states of the macroscopic measuring devices in and of themselves within quantum theory, that in turn means *we cannot even use ordinary language, save in the P-mode*, to describe the laboratory events as *our experiences*. Bohr increasingly came to recognize this, even if none too clearly, during his long career in interpreting quantum theory. Whereas he titled his 1927 set of essays on the physical content of quantum theory as "Atomic Theory and Description of Nature", he titled his subsequent set of essays in 1954 as "Atomic Physics and *Human Knowledge*" (emphasis mine). The shift in the wording emphasizes the point that, as long as we use only that range of the ordinary language which embodies classical common-sense notions, we can regard even laboratory observations as only our experiences. That is to say, in reporting such observations, ordinary language can operate *only* in the *P*-mode. Nothing more can be said about the predictions of quantum theory.[19] Any attempt to shift to the usual *R*-mode of ordinary language to speak of an underlying real state of even just the *macroscopic measuring instrument* is made logically inadmissible *within quantum* theory, due to the inseparability principle.

Let me spell out the reasoning behind the above conclusion in some detail. Let us say we have an observation experience, "the meter needle is pointing to +1." In a classical theory, such as electromagnetism, the observation experience might be interpreted as, say, the measurement of a current of 1 Ampere. Such an interpretation would emerge in the following sequence (as per the conception of the scientific realism given in Figure 4):

Stage 1: Treat the observation as our experience. Ordinary language is functioning in *P*-mode.

Stage 2: Treat the observation as an event in the external world. Ordinary language is functioning in the *R*-mode. There is a real meter in the world whose needle is pointing at +1.

Stage 3: The observation (of Stage 2) corresponds to measuring the magnitude of a physical quantity represented (say) by *I* in the theory. (Scientific observation using formal terms)

Stage 4: The observation (of Stage 3) is a measurement of *current*. (Scientific observation restated, with the formal terms named using ordinary language words).

Stage 5: The current of 1 Ampere was flowing in the circuit prior to the observation.

Let us say that in an experiment involving quantum theory, we have the *same* observation experience, "the meter needle is pointing to +1." Let us say the physicist interprets it as a measurement of the spin of an electron. That is a Stage 4 interpretation of the observation experience. As per the standard view, it is taken that in quantum theory too we go from Stage 1 to Stage 4 in a similar manner. It would be said that only Stage 5 would fail, i.e., the conclusion that the spin of the electron was +1 *prior* to the observation would be said to be invalid.

However, due to the 'inseparability hypothesis' as discussed above, Stages 2 through 5 *would all fail* in quantum theory. If so, given the observation experience, we cannot even talk about an observation-independent state of the *macroscopic* measuring device. In saying this we do not mean to deny the validity of the current common-sense realism in its own domain, i.e., at the level of everyday thinking, independent of science. For, independent of quantum theory, we *can* continue to interpret the observation experience as indicating the objective state of the needle of a real meter in the real world that existed even prior to our observation. However, if one wants to render the observation *experience* (of Stage 1) as a *scientific observation* <u>within quantum theory</u>, then due to (our reading of) the quantum inseparability, even Stage 2 of classical scientific realism fails. Reason: just as the state of the observed system cannot be separated from the observation, because the electron and the measuring device form a single epistemic whole, neither can the state of the measuring device be separated from the observation, *for the same reason*. Hence, we are (logically) prevented from using the observation experience to talk about a pre-existing real state for either the electron *or the measuring device*. I have called this the 'observation problem', to distinguish it from the usual measurement problem. I have argued elsewhere that this observation problem is logically prior to the measurement problem, and that a solution to the former could *avoid* the latter altogether (see Gomatam, 1999, 2003b).

---

[19] Thus Bohr: "The decisive point is that the physical content of quantum mechanics is *exhausted* by its power to formulate statistical laws governing observations obtained under conditions specified in plain language." [1963, p.12, emphasis mine]

Solving the 'observation problem' would evidently require us to revise our conception of the nature of a 'real' object at the *macroscopic* level so we can go from the *P*-mode of ordinary language to the *R*-mode of ordinary language at Stage 2 in a non-classical and quantum-compatible way. That is to say, a revision to our notions of reality would have to occur at the macroscopic level *before* we can hope to realistically interpret the quantum physical properties at the microscopic level at Stage 3 and Stage 4.

Of course, the current praxis of physicists is to shrug off the above pointed-out inseparability problem as it occurs at the level of interpreting the observation itself (at Stage 2), and simply assume the validity of the current classical *R*-mode of ordinary language to interpret the observation experience as indicating a pre-existing state of the macroscopic device (how else could we understand an observation experience of the sort "the meter needle pointing to +1"). Then they notice that Stage 5, however, fails in quantum theory (i.e., the observed state of the microscopic electron could not have existed prior to the observation) and *thus* physicists battle with the problem of measurement and hence the physical understanding of quantum theory.[20]

In our opinion, this approach is the result of viewing the current classical-physics-compatible common-sense thinking as unalterably given to us at the level of experience, and hence as the only mode of common-sense thinking possible. Even Bohr, despite his keen insight that quantum theory indicates an epistemic inseparability at the level of observation, held this restrictive attitude.

> As our knowledge becomes wider, we must always be prepared, therefore, to expect alterations in the points of view best suited for the ordering of our experience. In this connection we must remember, above all, that, as a matter

---

[20] In some sense, this point also formed the nub of Einstein's difficulties with the standard response to quantum theory: "The "macroscopic" and the "microscopic" are so inter related that it appears impracticable to give up this program in the "microscopic" alone." [Einstein, 1969, p.673]

> of course, *all new experience makes its appearance within the frame of our customary points of view and forms of perception*. [Bohr, 1927, p.1, emphasis mine]

However, it is not necessary to stay stuck in this box. We could conceivably try to identify an alternate range of common-sense concepts (and associated forms of perception) to describe the observation *qua* sense experiences, using ordinary language concepts that are more quantum-compatible (see Gomatam, 1999 in this regard). But even if we choose to insist that the classical range of common-sense object notions and their corresponding states are the only ones we can use to interpret the observations with respect to macroscopic measuring devices, the move to treat so simple a sense experience as "meter needle pointing to +1" as referring to an independent state of the measuring device requires *justification* in quantum theory due to inseparability of the quantum description of a system from the experimental context in which it is being observed. Pre-quantum theory could afford to assume naïve realism without justification. Whereas, as Bohr wrote:

> The quantum postulate implies that any observation of atomic phenomena will involve an interaction with the agency of observation not to be neglected. Accordingly, an independent reality in the ordinary physical sense can neither be ascribed to the phenomena *nor to the agencies of observation*.
> (Bohr 1934, 54, italics mine)

But the project of justifying our move from the *P*-mode to the usual *R*-mode of *ordinary language*, naïve realism, *is* the problem of philosophical realism, not scientific realism. Thus, physics and empirical philosophy stand closely connected in quantum theory, unlike in the era of classical physics.

However, contemporary philosophy has already fallen short of even the task of providing a justification for going over from *P*-mode to the *classical R*-mode of ordinary language. With quantum theory, the problem is more compounded, since quantum formalism disallows even the simple idea of 'position' in the current classical *R*-mode sense. If so, we are faced with nothing less than the task of finding a *quantum-compatible* version of *R*-mode thinking *in ordinary language,* and then justifying the transition from *P*-mode to this new *R*-mode, in order to realistically interpret the observation experience at Stage 2 in quantum theory.

Doing only philosophy *of* science, philosophers working in the foundations of quantum theory have so far remained satisfied with interpreting the measurement problem that the physicists have created in the first place by retaining the quasi-classical notion of 'position'.

Based on the discussion above, it is appropriate to ask whether it is possible to discover an alternative range of concepts in common-sense thinking regarding the states of macroscopic objects (and their properties) that could provide an alternative and quantum-compatible way of relating the *P*-mode of ordinary language to the *R*-mode in *common-sense thinking*, so that the physicists can consider interpreting the very observational content of quantum theory in a new light.

Any revision to common-sense notions about the states of macroscopic objects can also be expected to revise the very notion of matter we presently have at the level of common-sense thinking, identified by Descartes as *res extensa*.

## 10.0   On Extending the Cartesian 'cut'

Descartes identified just two ontological categories, extended matter and un-extended thinking substance. Their clean separation paved the way for the possibility of studying the workings of matter independent of the workings of the mind. Descartes saw such ontologically independent matter as being characterized primarily by its extension. That is to say, a stationary object has extension in the sense it has a continuous presence in an extended region. In common-sense thinking as well as in classical physics, this extended presence is averaged out as a single location, say, a center of gravity. In this sense, the property of being in an absolute location can be said to be fundamental to our (classical) conception of matter. Indeed, in all classical theories, absolute position is the most basic observation in physics.

We have seen that such a notion of absolute location, however, is incompatible with the quantum formalism. Indeed, in the current standard approach in quantum theory, the ontological quantum object is taken to not have a well-defined position. Rather, it is in a state of 'superposition', indicating a 'smeared out' *presence* in a localized region. We must note that this 'presence' is of a fundamentally non-classical character. It is *not* the case that the state of superposition means that the electron is simultaneously present in more than one location (in the classical absolute sense of the word location), as is often popularly, but mistakenly portrayed.  Rather, the quantum object —the electron or photon— has an extended smeared-out presence in a region that is non-classical, and which is formally characterized by complex-valued amplitudes whose physical meaning is not properly understood, even now, after Schrödinger developed the original formalism eight decades ago.

Tentatively, this objective presence is treated as a '*quantum* potentiality'. Nevertheless, physicists have managed to give (via a Born's rule) an *operational meaning* to this quantum potentiality that is entirely *classical*: the square of the complex-valued amplitude yields the probability for the quantum object to be found in an actual observation in *one* of many *classical locations*, i.e., one of many narrow regions. In this way, the classical notion of 'position' (as an absolute location in space and time) has returned as the basic observation in quantum physics too. Such a quasi-classical approach has brought success in practical applications of quantum formalism, and has allowed us to retain the Cartesian conception of a material object as having a single, absolute location *at the level of observation*; but equally, it has failed to bring clarity about the ontological quantum conception of matter itself.

It should now be easy to see why many attempts to clarify the quantum ontological situation have tried to trace a connection directly between quantum theory and consciousness.[21] Such attempts are a direct consequence of implicitly remaining within the Cartesian dualism. For, if the physical reality described by quantum mechanics at the ontological level could not be of the classical Cartesian extended substance, and the Cartesian system allows for only one other category, namely mind, it follows that the reality described by quantum mechanics must pertain to mind.

However, this might be too quick a jump. What if there are other kinds of *matter*, besides *res extensa*, which Descartes may have lumped together under the label of mind or 'consciousness'? In that case, what would seem indicated as a description of consciousness may well turn out to be a description of another level of matter that is in-between *res extensa* and consciousness. This may still not solve the 'consciousness-matter' interaction problem (or the mind-body problem), but at least it might open the way for an objective interpretation of matter that goes past the classical concept of matter without directly bringing in a role for a non-physical consciousness.

---

[21] See, among others, London, F. & Bauer, E. in J.A. Wheeler, W.H. Zurek, *Quantum Theory and Measurement* Princeton U.P.: Princeton, New Jersey, 1983; Wigner, E.P. (1961) "Remarks on the Mind-Body Question" in *The Scientist Speculates*, I.J. Good, Editor, W. Heinemann: London; Goswami, A. [1993] *The Self-Aware Universe*, Putnum Pub group:NY; Penrose, R. (1994) *Shadows of the Mind: A search for the missing science of consciousness* Oxford University Press: Oxford; Stapp, P. (1993) *Mind, Matter and Quantum Mechanics*, Springer-Verlag: Berlin, NY.

As per the foregoing analysis, the problem with the Cartesian 'cut' between mind and matter, or subject and object, would *not* be that quantum theory requires undoing such a cut, but rather the Cartesian cut has to be further *deepened*. In the sense, we can envisage other levels of matter besides *res extensa* that may have been swept under the common rubric of 'mind' in the simplistic Cartesian dualism.

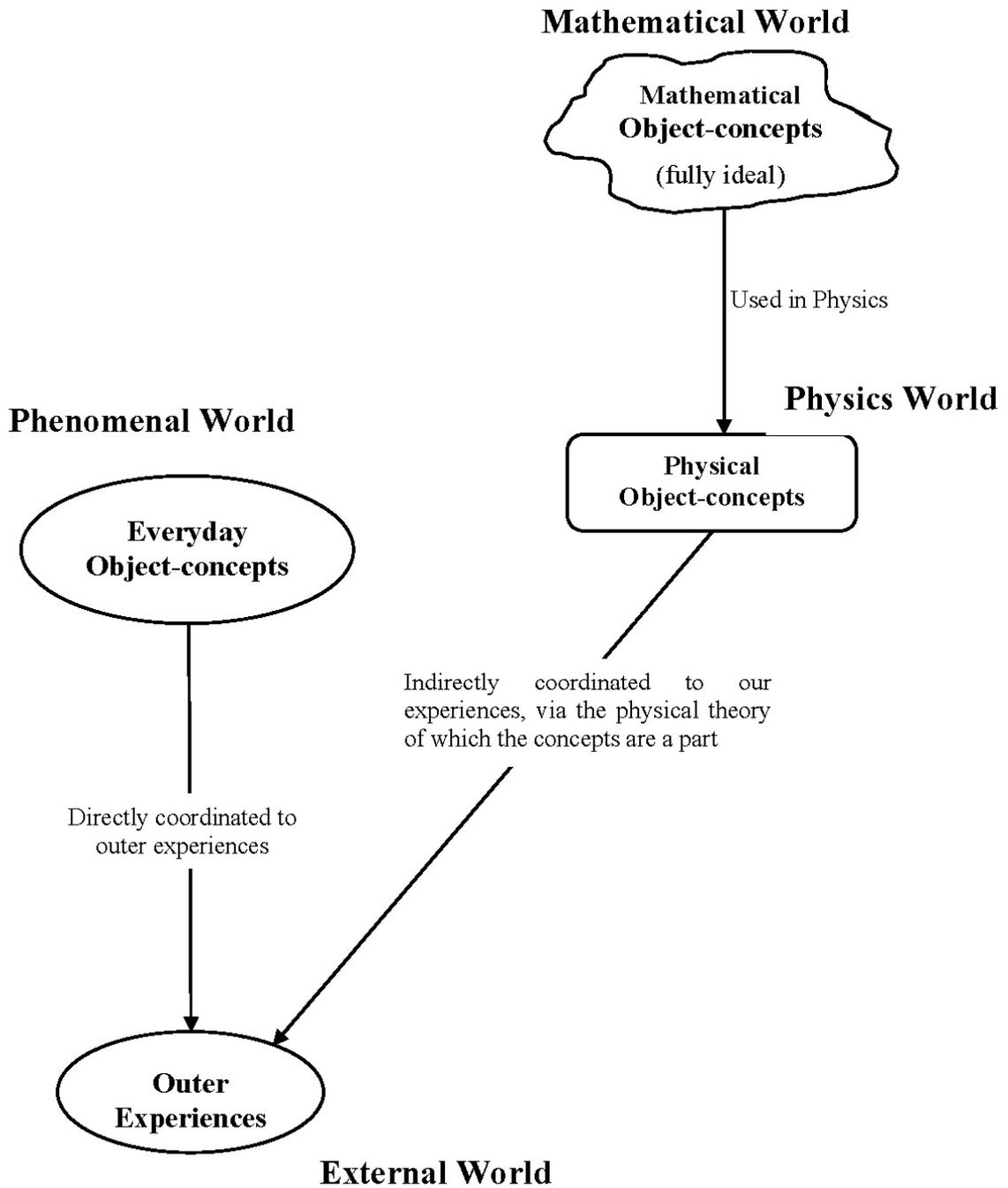

**Figure 1**

Different conceptual worlds and their inter-relationship

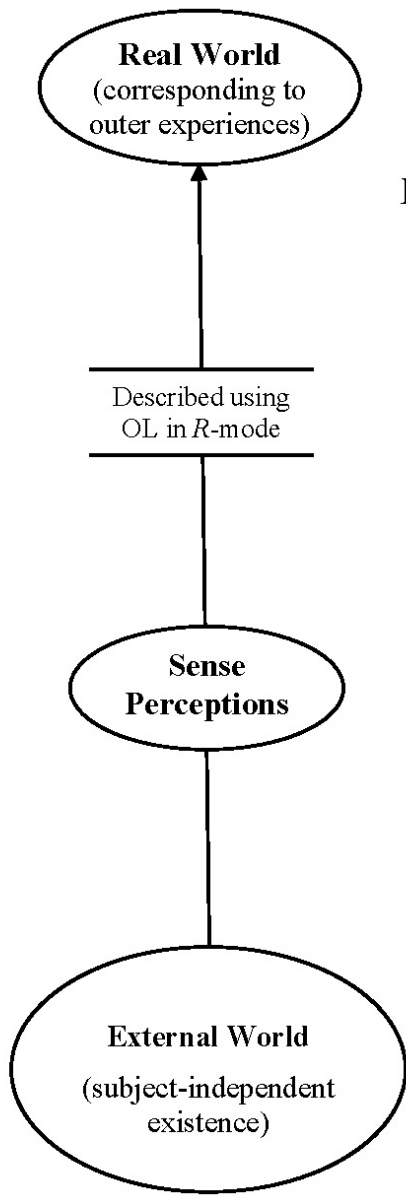

**Naïve or commonsense realism**

a) believes that we directly comprehend the world as it is,

b) that ordinary language (OL) can and must passively express this comprehension, and

c) thus equates the known, or knowable real world with the external world

Legend: $R$-mode = Realist mode

## Figure 2
Naïve realism

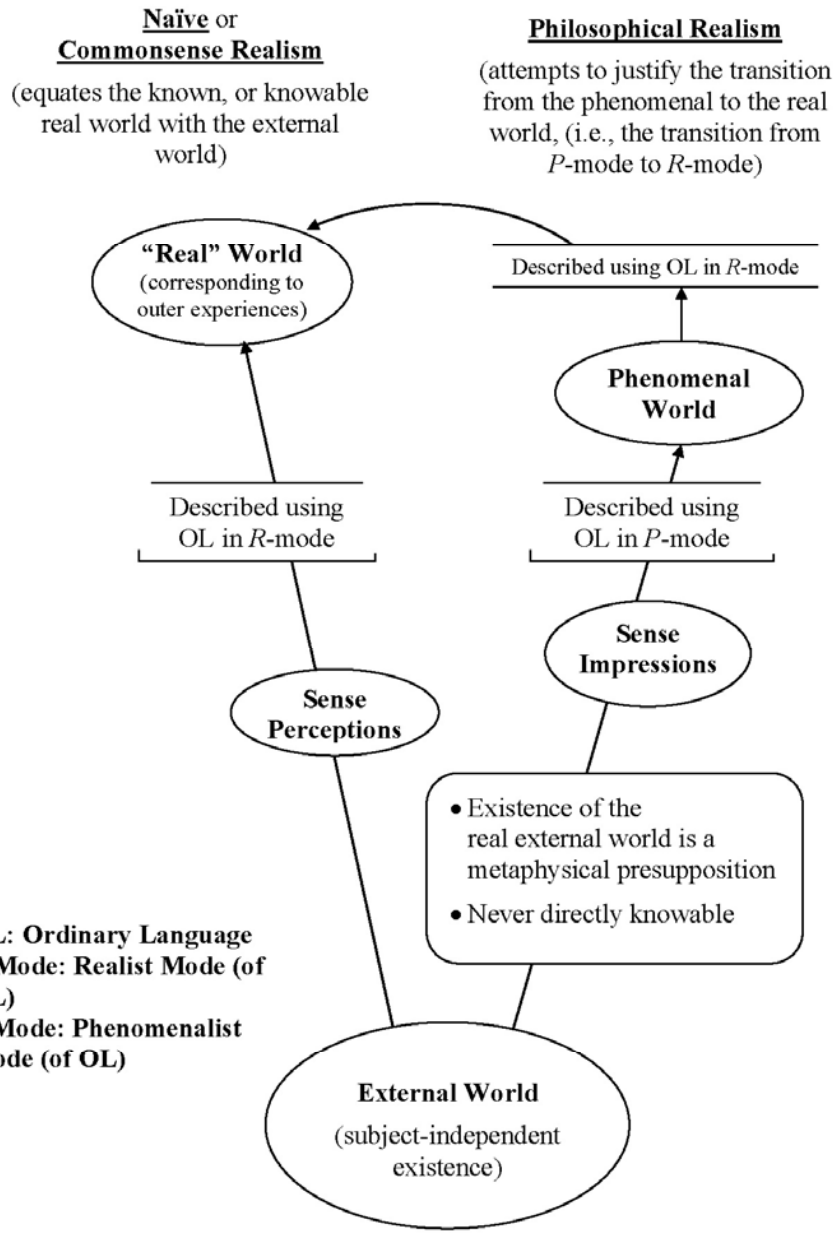

Figure 3

Naïve and philosophical realisms

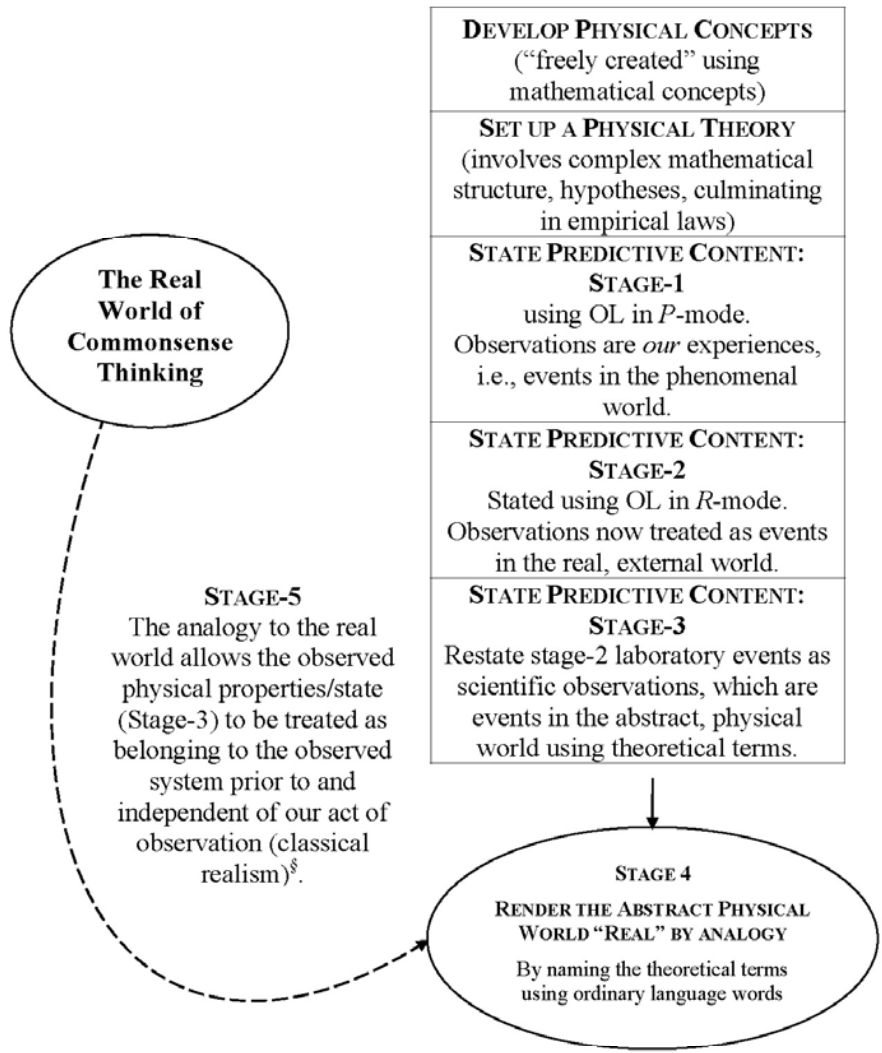

§In the orthodox or textbook view of quantum theory, classical realism is taken to fail in Stage 5. The other stages are taken to remain valid. In section 9.0 of this paper, we make the proposal that Stage-2 through Stage 5 might fail in quantum theory.

**Figure 4**
Classical Scientific realism

**Appendix A**

We present a simple technical argument [due to Peres, 1984, p. 644] as to why quantum theory seems to entail the view that measurement creates what is observed.

The total angular momentum L for a particle is given by

$$L^2 = l(l+1)\hbar \text{ where } l = \tfrac{1}{2}, \tfrac{3}{2}, \tfrac{5}{2}... \quad [1]$$

The angular momentum component $L_i$ in any chosen direction x, y or z ($L_x$, $L_y$ or $L_z$) is given by the formula

$$L_i = m\hbar \text{ where } -l < m < +l \quad [2]$$

Suppose we performed a measurement corresponding to the total angular momentum observable L and obtained result corresponding to $l = \tfrac{3}{2}$. The total angular momentum $L^2$ would then have the value $\tfrac{15}{4}$ (by eq. [1]).

If we measure one of the angular momentum components, say $L_x$, the result will always turn out to be $\pm\tfrac{1}{2}$ or $\pm\tfrac{3}{2}$ (by eq. [2]). Can we say that $L_x$ really had one of these values before the measurement, but we just did not know which one? In that case, $L_x^2$ would *always* have objectively one of the values, $\tfrac{1}{4}$ or $\tfrac{9}{4}$, and likewise $L_y^2$ and $L_z^2$. However, it is impossible to combine the values $\tfrac{1}{4}$ and $\tfrac{9}{4}$ to fulfill $L_x^2 + L_y^2 + L_z^2 = \tfrac{15}{4}$.

We are therefore led to the conclusion that $L_x$ is forced into one of the two values, $\pm\tfrac{1}{2}, \pm\tfrac{3}{2}$ by the act of measurement; that these values did not exist in some 'objective reality' prior to the measurement, but are created by it.